\input ./style/arxiv-general.cfg
\documentclass[seceqn,MSNbibl,nameyear,dvips]{arxstspdf}
\makeatletter
   \@ifpackageloaded{graphicx}{}{\usepackage{graphicx}}
\makeatother
\usepackage{multirow}
\usepackage{flushend}
\usepackage{stfloats}

\volume{30}
\issue{2}
\pubyear{2015}
\firstpage{199}
\lastpage{215}
\doi{10.1214/14-STS504}
\docsubty{FLA}

\makeatletter

\setattribute{ead}{sep}{; }
\newcommand{\eqref}[1]{(\ref{#1})}
\newcommand{\Ex}{\mathsf{E}}

\makeatother

\begin{document}
\begin{frontmatter}

\title{Multi-armed Bandit Models for the Optimal Design of Clinical
Trials: Benefits and Challenges}
\runtitle{MAB Models for the Optimal Design of Clinical Trials}

\begin{aug}
\author[A]{\fnms{Sof\'ia S.}~\snm{Villar}\corref{}\ead[label=e1]{sofia.villar@mrc-bsu.cam.ac.uk}},
\author[A]{\fnms{Jack}~\snm{Bowden}\ead[label=e2]{jack.bowden@mrc-bsu.cam.ac.uk}}
\and
\author[A]{\fnms{James}~\snm{Wason}\ead[label=e3]{james.wason@mrc-bsu.cam.ac.uk}}
\runauthor{S. S. Villar, J. Bowden and J. Wason}

\affiliation{MRC Biostatistics Unit, Cambridge and Lancaster University}

\address[A]{Sof\'ia S. Villar is
Investigator Statistician at MRC BSU,
Cambridge and Biometrika post-doctoral research fellow,
Jack Bowden
is
Senior Investigator Statistician at MRC BSU, Cambridge
 and  James
Wason is Senior Investigator Statistician at MRC BSU, Cambridge, MRC Biostatistics Unit,
Cambridge Institute of Public Health,
Forvie Site, Robinson Way,
Cambridge Biomedical Campus,
Cambridge CB2 0SR,
United Kingdom \printead
{e1,e2,e3}.}
\end{aug}

%
\begin{abstract}
\emph{Multi-armed bandit} problems (MABPs) are a special type of
optimal control problem well suited to model resource allocation under
uncertainty in a wide variety of contexts. Since the first publication
of the optimal solution of the \emph{classic} MABP by a dynamic
index rule, the bandit literature quickly diversified and emerged as an
active research topic.
Across this literature, the use of bandit models to optimally design
clinical trials became a
typical motivating application, yet little of the resulting theory has
ever been used in the actual design and analysis of clinical trials.
To this end, we review two MABP decision-theoretic approaches to the
optimal allocation of treatments in a clinical trial:
the infinite-horizon Bayesian Bernoulli MABP and the finite-horizon
variant. These models possess distinct theoretical properties and lead
to separate allocation rules in a clinical trial design context.
We evaluate their performance compared to other allocation rules,
including fixed randomization.
Our results indicate that bandit approaches offer significant
advantages, in terms of assigning more patients to better treatments,
and severe limitations, in terms of their resulting statistical power.
We propose a novel bandit-based patient allocation rule that overcomes
the issue of low power, thus removing a potential barrier for their use
in practice.
\end{abstract}

%
\end{frontmatter}

\section{Introduction} \label{sec:intro}

Randomized controlled trials have become the gold-standard approach in
clinical research over the last 60 years. Fixing the probability of
being assigned to each arm for its duration, it removes
(asymptotically) any systematic differences between patients on
different arms with respect to all known or unknown confounders. The
frequentist operating characteristics of the standard approach (e.g.,
the type-I error rate and power) are well understood, and the size of
the trial can easily be chosen in advance to fix these at any level the
practitioner desires. However, while it is important for a clinical
trial to be adequately powered to detect a significant difference at
its conclusion, the well-being of patients during the study itself must
not be forgotten.

MABPs are an idealized mathematical decision framework for deciding how
to optimally allocate a resource among a number of competing uses,
given that such allocation is to be done \emph{sequentially} and under
\emph{randomly evolving conditions}. In its simplest version, the
resource is work which can further be devoted to only one use at a
time. The uses are treated as independent ``projects'' with a binary
outcome which develop following Markov rules. Their roots can be traced
back to work produced by \citet{thompson1933likelihood}, which was
later continued and developed in \citet{robbins1952some}, \citet{bellman1956problem},
and finally \citet{gijo74}. Although their scope
is much more general, the most common scenario chosen to motivate this
methodology is that of a clinical trial which has the aim of balancing
two separate goals:
\begin{itemize}
\item To correctly identify the best treatment (\emph{exploration or
learning}).
\item To treat patients as effectively as possible during the trial
(\emph{exploitation or earning}).
\end{itemize}
One might think that these two goals are naturally complementary, but
this is not the case.
Correctly identifying the best treatment requires some patients to be
assigned to all treatments, and therefore the former acts to limit the latter.

Despite this apparent near-perfect fit between a real-world problem and
a mathematical theory, the MABP has yet to be applied to an actual
clinical trial. Such a state of affairs was pointed out early on by
Peter Armitage in a paper reflecting upon the use in practice of
theoretical models to derive optimal solutions for problems in clinical trials:

\begin{quote}
Either the theoreticians have got hold of the wrong problem, or the
practising triallists have shown a
culpable lack of awareness of relevant theoretical developments, or
both. In any case, the
situation does not reflect particularly well on the statistical
community  (\citeauthor{armitage1985search}, \citeyear{armitage1985search}, page 15).
\end{quote}

A very similar picture is described two decades later in \citet
{palmer2002ethics} when discussing and advocating for the use of
``learn-as-you-go'' designs as a means of alleviating many problems
faced by those involved with clinical trials today. More recently, Don
Berry---a leading proponent of the use of Bayesian methods to develop
innovative adaptive clinical trials---also highlighted the resistance
to the use of bandit theoretical results:

\begin{quote}
But if you want to actually use the result then people will attack your
assumptions. Bandit problems are good examples. An explicit assumption
is the
goal to treat patients effectively, in the trial as well as out. That
is controversial (\,\ldots) \citep{stangl2012celebrating}.
\end{quote}

In view of this, a broad goal of this article is to contribute to
setting the ground for change by reviewing a concrete area of
theoretical bandit results, in order to facilitate their application
in practice. The layout of the paper is as follows: In
Section~\ref{sec:2 mabp} we first recount the basic elements of
the Bayesian Bernoulli MABP. In Section~\ref{classic MABP} we focus
on the infinite-horizon case, presenting its solution in terms of
an index rule---whose optimality was first proved by Gittins and
Jones over 30 years ago. In Section~\ref{restless MABP} we review the
finite horizon variant by reformulating it as an equivalent
infinite-horizon restless MABP, which further provides a means to
compute the index rule for the original problem. In Section~\ref{suitability} we compare, via simulation, the performance of the MABP
approaches to existing methods of response adaptive allocation
(including standard randomization) in several clinical trial settings.
These results motivate the proposal of a composite method, that
combines bandit-based allocation for the experimental treatment arms
with standard randomization for the control arm. We conclude
in Section~\ref{Conclusion} with a discussion of the existing barriers to
the implementation of bandit-based rules for the design of clinical
trials and point to future research.

\section{The Bayesian Bernoulli Multi-armed Bandit Problem}\label
{sec:2 mabp}

The Bayesian Bernoulli $K$-armed bandit problem
corresponds to a MABP in which
only one arm can be worked on at a time $t$, and work on arm $k= 1,
\dots, K$ represents
drawing a sample observation 
from a Bernoulli population $Y_{k,t}$ with unknown parameter $p_k$,
``earning'' the observed value $y_{k,t}$ as a reward (i.e., either $0$
or $1$).
In a clinical trial context, each arm represents a treatment with an
unknown success rate. 
The Bayesian feature is introduced by letting each parameter $p_k$ have
a Beta prior with parameters $s_{k,0}$ and $f_{k,0}$ such that
$(s_{k,0},f_{k,0}) \in\mathbb N^2_{+}$ before the first sample
observation is drawn (i.e., at $t=0$).
After having observed $S_{k,t}=s_{k,t}$ successes and $F_{k,t}=f_{k,t}$
failures, with $(S_{k,t},F_{k,t}) \in\mathbb N^2_{0}$ for any $t\ge1$,
the posterior density is a Beta distribution with parameters
$(s_{k,0}+s_{k,t},f_{k,0}+f_{k,t})$.

Formally, the Bernoulli Bayesian MABP is defined by letting each arm
$k$ be a discrete-time Markov Control Process (MCP) with 
the following elements:
\begin{longlist}[(b)]
\item[(a)] The \emph{state space}: 
$\mathbb{X}_{k,t}= \{(s_{k,0}+S_{k,t},f_{k,0}+F_{k,t})\in\mathbb
N_+^2: S_{k,t}+F_{k,t} \le t,  \mbox{ for } t=0,1,\dots, T \}$ 
which
represents all the possible 
two-dimensional vectors of information on the unknown parameter $p_k$
at time $t$. We denote the available information on treatment $k$ at
time $t$ as $\mathbf{x}_{k,t}=(s_{k,0}+S_{k,t}, f_{k,0} +F_{k,t})$ and
the initial prior as $\mathbf{x}_{k,0}=( s_{k,0},f_{k,0})$. In a
clinical trial context, the random vector $(S_{k,t},F_{k,t})$
represents the number of successful and unsuccessful patient outcomes
(e.g., response to treatment, remission of tumor, etc.).

\item[(b)] The \emph{action set} $\mathbb A_k$ is a binary set
representing the action of drawing a sample observation from population
$k$ at time $t$ ($a_{k,t}=1$) or not ($a_{k,t}=0$). In a clinical
context, the action variable stands for the choice of assigning patient
$t$ to treatment arm $k$ or not.

\item[(c)] The Markovian \emph{transition law} $\mathcal{P}_k(\mathbf
{x}_{k,t+1}|\mathbf{x}_{k,t},\break a_k)$ describing the evolution of the
information state variable in population $k$ from time $t$ to $t+1$ is
given~by
%
\begin{equation}
\label{dynamics1}\quad \mathbf{x}_{k,t+1} = \cases{ (s_{k,0}+s_{k,t}+1
, f_{k,0}+f_{k,t} ), \cr
\quad
 \mbox{if }  a_{k,t}=1 \cr\quad\mbox{w.p. } \displaystyle\frac{ s_{k,0}+
s_{k,t}}{s_{k,0}+f_{k,0}+s_{k,t}+f_{k,t}}  ,
\cr
(s_{k,0}+s_{k,t},f_{k,0}+f_{k,t}+1
), \cr
\quad\mbox{if  } a_{k,t}=1 \cr
\quad\mbox{w.p. }\displaystyle \frac
{f_{k,0}+f_{k,t}}{s_{k,0}+f_{k,0}+ s_{k,t}+f_{k,t}} ,
\cr
\mathbf{x}_{k,t}, \cr
\quad\mbox{if } a_{k,t}=0\mbox{ w.p.
 }1, }
\end{equation}
for any $\mathbf{x}_{k,t} \in\mathbb{X}_{k,t}$ and where
w.p. stands for
``with probability.''

\item[(d)] The expected rewards and resource consumption functions are
%
\begin{eqnarray}
\label{rewards1}\quad \mathcal R(\mathbf{x}_{k,t}, a_{k,t})&=&
\frac{s_{k,0}+s_{k,t}}{s_{k,0}+f_{k,0}+s_{k,t}+f_{k,t}}a_{k,t} ,
\nonumber\\[-10pt]\\[-10pt]
\mathcal C(\mathbf{x}_{k,t},
a_{k,t})
&=& a_{k,t},\nonumber
\end{eqnarray}
for $t =0,1, \ldots, T-1$,
where, in accordance to \eqref{dynamics1},
a reward (i.e., a treatment success) in arm $k$ arises only if that arm
is worked on and with a probability given by the posterior predictive
mean of $p_k$ at time $t$
and resource consumption is restricted by the fact that (at most) one
treatment can be allocated to every patient in the trial, that is, $\sum_{k=1}^{K}a_{k,t}\le1$ for all $t$.
\end{longlist}

A rule is required to operate the resulting MCP, indicating which
action to take for each of the $K$ arms, for every possible combination
of information states
and at every time $t$, until the final horizon $T$.
Such a rule forms a sequence of actions $\{a_{k,t}\}$, which depends on
the information available up to time $t$, that is, on $\{\mathbf
{x}_{k,t}\}$, and it is known as a \emph{policy} within the Markov
Decision Processes literature.
To complete the specification of this multi-armed bandit model as an
\emph{optimal control model},
the problem's \emph{objective function} must be selected. Given an
objective function and a time horizon,
a multi-armed bandit optimal control problem is mathematically
summarized as the problem of finding a feasible policy, $\pi$, in $ \Pi
$ (the set of all the feasible policies given the resource constraint)
that optimizes the selected performance objective.

The performance objective in the Bayesian\break Bernoulli MABP is to maximize
the Expected Total Discounted (ETD) number of successes after $T$
observations, letting $0 \le d < 1$ be the discount factor. Then, the
corresponding bandit optimization problem is to find a discount-optimal
policy such that
%
\begin{eqnarray}
\label{eq:dop_1} &&\quad V_D^*(\mathbf{\tilde{x}_0})\nonumber
\\
&&\quad\quad= \max
_{{\pi} \in{{\Pi}} } \Ex ^{{\pi}} \Biggl[\sum
_{t = 0}^{T-1} \sum_{k=1}^K
d^t \frac
{s_{k,0}+S_{k,t}}{s_{k,0}+f_{k,0}+S_{k,t}+F_{k,t}} \\
&&\quad\hphantom{\quad= \max
_{{\pi} \in{{\Pi}} } \Ex ^{{\pi}} \Biggl[\hspace*{62pt}}{}\cdot a_{k,t}\Big|\mathbf {\tilde{x}_0}=
(\mathbf{x}_{k,0} )_{k=1}^{K} \Biggr],\nonumber
\end{eqnarray}
where $\tilde{\mathbf{x}}_0$
is the initial joint state, $\Ex
^{{\pi}}[\cdot]$ denotes
expectation under policy ${\pi}$ and transition probability rule \eqref
{dynamics1},
$V_D^*(\tilde{\mathbf{x}}_0)$ is the optimal expected total discounted
value function conditional on the initial joint state being equal to
$\mathbf{\tilde{x}_0}$ (for any possible joint initial state),
and where, given the resource constraint, the family of admissible
feasible policies $\Pi$ contains the sampling rules $\pi$ for which it
holds that
$\sum_{k=1}^{K}a_{k,t}\le1$ for all $t$.

A generic MABP formally consists of $K$ discrete-time MCPs with their
elements defined in more generality, that is, (a)~the \emph{state
space}: a Borel space, (b)~the binary \emph{action set}, (c)~the \emph
{Markovian transition law}: a stochastic kernel on the state space
given each action and~(d) a \emph{reward} function and a \emph{work
consumption} function: two measurable functions. As before, the MABP is
to find a policy that optimizes a given performance criterion, for
example, it maximizes the ETD net rewards.

\citet{robbins1952some} proposed an alternative version of the Bayesian
Bernoulli MABP problem
by considering the
average \emph{regret} after allocating $T$ sample observations [for a
large $T$ and for any given and unknown $(p_k)_{k=1}^K$].
For the Bayesian Bernoulli MABP, the total regret $\rho$ is defined as
%
\begin{eqnarray}
\label{regret}&& \rho= T \max_{k} \{p_k\}-
\Ex^{\pi} \Biggl[ \sum_{t=0}^{T-1}
\sum_{k=1}^{K}a_{k,t}
Y_{k,t} \Biggr] \nonumber\\[-10pt]\\[-10pt]
&&\quad\mbox{for some } (p_k)_{k=1}^K.\nonumber
\end{eqnarray}

A form of \emph{asymptotic optimality} can be defined for sampling
rules $\pi$ in terms of \eqref{regret}
if it holds that for any $(p_k)_{k=1}^K$,
$\lim_{T \to\infty}\frac{\rho}{T}=0$.
A necessary condition for a rule to attain this property is to sample
each of the $K$ populations infinitely often, that is, to continue to
sample from (possibly) suboptimal arms for every $t < \infty$.
In other words, asymptotically optimal rules have a strictly positive
probability of allocating  a patient to every arm at any point of the
trial. Of course, within the set of asymptotically optimal policies
secondary criteria may be defined and considered
(see, e.g., \citeauthor{lai1985asymptotically}, \citeyear{lai1985asymptotically}).
As it will be illustrated in Section~\ref{suitability}, objectives in
terms of \eqref{eq:dop_1} or \eqref{regret} give rise to sampling rules
with distinct statistical properties. Asymptotically optimal rules,
that is, in terms of \eqref{regret}, maximize the \emph{learning} about
the best treatment, provided it exists, while the rules that are
optimal in terms of \eqref{eq:dop_1} maximize the mean number of total
successes in the trial.

\section{The Infinite-horizon Case: A Classic MABP} \label
{classic MABP}

We now review the solution giving the optimal policy to optimization
problem (\ref{eq:dop_1}) in the infinite-horizon setting by letting
$T=\infty$. In general, as MABPs are a special class of MCPs,
the traditional technique to address them is via a dynamic programming
(DP) approach. Thus, the solution to \eqref{eq:dop_1}, according to
Bellman's principle of optimality \citep{bellman1952theory}, is such
that for every $t=0, 1, \ldots$ the following DP equation holds:
%
\begin{eqnarray}
\label{eq:dp_1} 
&&V_D^*(\mathbf{x}_{1,t}, \dots,
\mathbf{x}_{K,t})\nonumber
\\
&&\quad =\max_{k } \biggl\{
\frac{s_{k,0}+s_{k,t}}{s_{k,0}+f_{k,0}+s_{k,t}+f_{k,t}}\nonumber
\\
&&\hphantom{=\max_{k } \biggl\{}\quad{}+ d \biggl( \frac
{s_{k,0}+s_{k,t}}{s_{k,0}+f_{k,0}+s_{k,t}+f_{k,t}}\nonumber
\\[-10pt]\\[-10pt]
&&\hphantom{=\max_{k } \biggl\{{}+ d \biggl(}\quad{}\cdot V_D^*(\mathbf {x}_{1,t},
\mathbf{x}_{k,t}+\mathbf{e}_1
, \dots,
\mathbf{x}_{K,t}) \nonumber
\\
&&\hphantom{=\max_{k } \biggl\{{}+ d \biggl(}\quad{} +\frac{f_{k,0}+f_{k,t}}{s_{k,0}+f_{k,0}+
s_{k,t}+f_{k,t}}\nonumber
\\
&&\hphantom{=\max_{k } \biggl\{{}+ d \biggl(+\,}\quad{}\cdot V_D^*(\mathbf{x}_{1,t},
\mathbf{x}_{k,t}+\mathbf{e}_2 
, \dots,
\mathbf{x}_{K,t}) \biggr) \biggr\}, \nonumber
\end{eqnarray}
where $\mathbf{e}_1$, $\mathbf{e}_2$ respectively denote the unit
vectors $(1,0)$ and $(0,1)$. Under the assumptions defining the
Bayesian Bernoulli MABP, the theory for discounted MCPs ensures the
existence of an optimal solution to \eqref{eq:dp_1}
and also the monotone convergence of the value functions $V_D^*(\tilde
{\mathbf{x}}_{t})$.
Therefore,
equation \eqref{eq:dp_1} can be approximately solved iteratively using
a backward induction algorithm.
%
\begin{figure}

\includegraphics{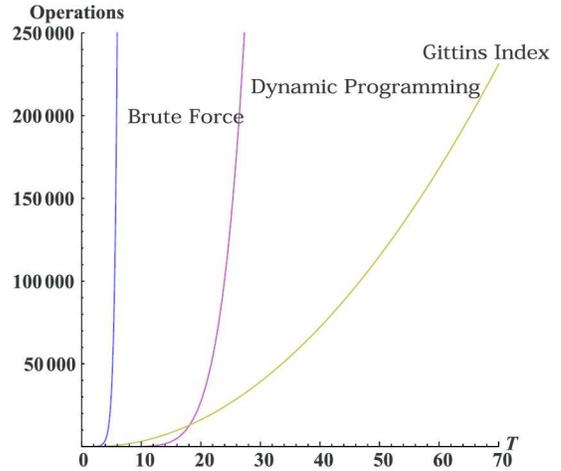}

\caption{The number of individual computations for an approximation to
the optimal rule in a particular instance of the Bayesian Bernoulli
MABP as a function of $T$ with $K=3$ and $d=0.9$ for the Brute force,
DP and Gittins index approaches.}
\label{fig:1}
\end{figure}

Unfortunately, as shown in Figure~\ref{fig:1}, such a DP technique suffers
from a severe computational burden, which is particularly well
illustrated in the \emph{classic} MABP where the size of the state
space grows with the truncation horizon $T$. To illustrate this fact,
consider the case of $K$ treatments with an initial uniform prior
distribution (i.e., $s_{k,0} = f_{k,0} = 1  \ \forall k$) and
truncation horizon to initialize the algorithm equal to $T$. The total
number of individual calculations [i.e., the number of successive
evaluations of $ V_D^*(\mathbf{x}_{1,t}, \dots, \mathbf{x}_{K,t})$]
required to find an approximate optimal solution by means of the DP
algorithm equals $\frac{(T-1)!}{(2K)!(T-2K-1)!}$. The precision of such
an approximation depends on $d$, for example, if $d\le0.9$ values to
four-figure accuracy are calculated for $T \ge100$. Therefore,
considering the problem with $K=3$ and $d=0.9$ (and hence $T\ge100$)
makes the intractability of the problem's optimal policy become
evident. (For a more detailed discussion see the \hyperref[index computation]{Appendix}.)

\subsection{The Gittins Index Theorem}\label{GItheorem}

The computational cost of the DP algorithm to solve equation \eqref
{eq:dp_1} is significantly smaller than the cost of a complete
enumeration the set of feasible policies $\Pi$ (i.e., the \emph{brute
force} strategy), yet it is still not enough to make the solution of
the problem applicable for most real world scenarios, with more than 2
treatment arms. For this reason the problem gained the reputation of
being extremely hard to solve soon after being formulated for the first
time, becoming a paradigmatic problem to describe the exploration
versus exploitation dilemma characteristic of any \emph{data-based
learning} process.

Such a state of affairs explains why the solution first obtained by
\citet{gijo74} constitutes such a landmark event in the bandit literature.
The Index theorem states that
if problem $P$ is an infinite-horizon MABP
with each of its $K$ composing MCPs having (1)~a finite action set
$\mathbb{A}_k$, (2) a finite or infinite numerable state space $\mathbb{X}_{k}$,
(3) a  Markovian transition law under the passive action $a_{k,t}=0$
(i.e., the \emph{passive} dynamics) such that
%
\begin{eqnarray}
\label{classic dyn} \mathcal P_k\bigl(x_k'|x_k,0
\bigr)&=& \mathcal{P}_k\bigl\{X_{k,t+1}=x_k'|
X_{k,t}=x_{k}, a_{k,t}=0\bigr\}\nonumber
\\[-5pt]\\[-10pt]
&=& 1 _{\{x_{k'}=x_k\}},\nonumber
\end{eqnarray}
for any $x_{k}, x_{k}'\in\mathbb{X}_k$,
where
$1_{\{x_{k'}=x_k\}}$ is an indicator variable for the event that the
state variable value at time $t+1$: $x_{k'}$ equals the state variable
value of state $t$: $x_{k}$, and (4) the set of feasible polices $\Pi$
contains all polices $\pi$ such that for all $t$
%
\begin{equation}
\label{reso dyn}\sum_{k=1}^{K}a_{k,t}
\le1,
\end{equation}
then there exists a real-valued index function $\mathcal G(x_{k,t})$,
which recovers the optimal solution to such a MABP when the objective
function is defined under a ETD criterion, as in \eqref{eq:dop_1}. Such
a function is defined as follows:
%
\begin{equation}
\label{Gittins} \hspace*{10pt}\mathcal G_k(x_{k,t})= \sup
_{\tau\ge1}\frac{\Ex
_{X_{k,t}=x_{k,t}} \sum_{i=0}^{\tau-1} \mathcal
R(X_{k,t+i},1) d^i }{\Ex_{X_{k,t}=x_{k,t}}  \sum_{i=0}^{\tau-1}
\mathcal C(X_{k,t+i},1) d^i },\hspace*{-20pt}
\end{equation}
where the expectation is computed with respect to the corresponding
Markovian (\emph{active}) transition law $\mathcal{P}_k(x_k'|x_k,1)$,
and $\tau$ is a stopping time. Specifically, the optimal policy $\pi^*$
for problem $P$ is to work on the bandit process with the highest index
value, breaking ties randomly. Note that the stopping time $\tau$ is
past-measurable, that is, it is based on the information available at
each decision stage only. Observe also that the index is defined as the
ratio of the ETD reward up to $\tau$ active steps
to the ETD cost up to $\tau$ active steps.

MABPs whose dynamics are restricted as in \eqref{classic dyn} (namely,
those in which passive projects remain frozen in their states) are
referred to in the specialized literature as \emph{classic} MABPs and
the name Gittins index is used for the function \eqref{Gittins}.
The Index theorem's significant impact derives from the possibility of
using such a result
to break the curse of dimensionality by decomposing the optimal
solution to a $K$-armed MABP in terms of its independent parts, which
are remarkably more tractable than the original
problem as shown in Figure~\ref{fig:1}. The number of individual
calculations required to solve problem \eqref{eq:dp_1} using the Index
theorem is of order $\frac{1}{2} (T-1) (T-2)$, which no longer explodes
with the truncation horizon $T$. Further, it is completely independent
of $K$, which means that a single index table suffices for all possible
trials, therefore reducing the computing requirements appreciably. (For
more details, see the \hyperref[index computation]{Appendix}.)

Such computational savings are particularly well illustrated in the
Bayesian Bernoulli  MABP where the Gittins index \eqref{Gittins} is
given by
\begin{equation}\label{GittinsB}
\mathcal G_k(\mathbf{x}_{k,t})= \sup_{\tau\ge 1}
\frac{\Ex_{\bolds{\cdot}} \sum_{i=0}^{\tau-1}\frac{s_{k,0}+S_{k,t+i}}{s_{k,0}+ f_{k,0}+S_{k,t+i}+ F_{k,t+i}} d^i }
{\Ex_{\bolds{\cdot}}  \sum_{i=0}^{\tau-1}d^i },\hspace*{-30pt}
\end{equation}
where\vspace*{1pt} $ \Ex_{\bolds{\cdot}}= \Ex_{\mathbf{X}_{k,t}=(s_{k,0}+s_{k,t}, f_{k,0}+f_{k,t})}$.
%

Calculations of the indices \eqref{GittinsB}
have been reported in brief tables as in \citet{gi79} and \citet
{robinson1982algorithms}. Improvements to the efficiency of this
computing the index have since been proposed by
\citet
{katehakis1985multi},
\citet
{katehakis1986computing}. Moreover, since the
publication of Gittins' first proof of the optimality result of the
index policy for a classic MABP in \citet{gijo74}, there have been
alternative proofs, each offering complementary insights and
interpretations. Among them, the proofs by \citet{whittle1980multi},
\citet{varaiya1985extensions}, \citet{weber1992gittins} and \citet
{bertsimas1996conservation} stand out.

\begin{table}
\caption{The (approximate) Gittins index values for an information
vector of $s_{0}+s_{t}$ successes and $f_0+f_t$ failures, where
$d=0.99$ and $T$ is truncated at $T=750$}
\label{tab:Gittinstable}
\begin{tabular*}{\columnwidth}{@{\extracolsep{\fill}}lcccccc@{}}
\hline
$\bolds{f/s}$& \textbf{1} &
\textbf{2} & \textbf{3} & \textbf{4} & \textbf{5} & \textbf{6} \\
\hline
{1} & $0.8699$ & $0.9102$ & $0.9285$ & $0.9395$& $0.9470$ & $0.9525$ \\
{2} & $0.7005$ & $\underline{0.7844}$ & $0.8268$ & $0.8533$ & $0.8719$&
$0.8857$ \\[1pt]
{3} & $0.5671$ & $\underline{\underline{0.6726}}$ & $0.7308$ & $0.7696$
&$0.7973$ &$0.8184$\\[3pt]
{4} & $0.4701$ &$0.5806$ & $0.6490$ & $\underline{0.6952}$ & $0.7295$
&$0.7561$ \\[1pt]
{5} & $0.3969$ & $0.5093$ & $0.5798$ &$ 0.6311$ & $0.6697$ &$0.6998$ \\
{6} & $0.3415$ & $0.4509$& $0.5225$ &$0.5756$ & $0.6172$ & $\underline
{\underline{0.6504}}$ \\
\hline
\end{tabular*}
\end{table}

\begin{figure}[b]

\includegraphics{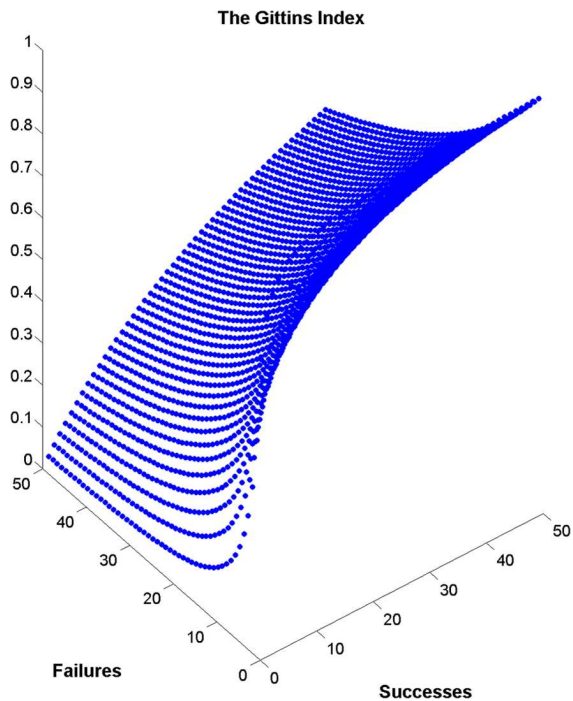}

\caption{The (approximate) Gittins index values for an information
vector of $s_{0}+s_{t}$ successes and $f_0+f_t$ failures, where
$d=0.99$ and $T$ is truncated at $T=750$.}
\label{fig:2}
\end{figure}

To elaborate a little more on the use of the Gittins index for solving
a $K$-armed Bayesian Bernoulli MABP in a clinical trial context, we
have included some values of the Gittins index in Table~\ref{tab:Gittinstable} and Figure~\ref{fig:2}. These values correspond to a
particular instance in which the initial prior for every arm is
uniform, the discount factor is $d=0.99$, the index precision is of 4
digits and we have truncated the search of the best stopping time to
$T=750$. The choice of $d=0.99$ is a widely used value in the related
bandit literature. In our example, since $0.99^{750}<10^{-3}$, patients
treated after this time
yield an almost zero expected discounted reward and are hence ignored.

The Gittins index policy assigns a number to every treatment (from an
extended version of Table~\ref{tab:Gittinstable}) based on the values of
$s_{k,t}$ and $f_{k,t}$ observed, and then prioritizes sampling the one
with the highest value. Thus, provided that we adjust for each
treatment prior, 
the same table can be used for making the allocation decision of all
treatments in a trial. Furthermore, the number of treatments need not
be prespecified in advance and new treatments may be seamlessly
introduced part way through the trial as well
(see \citeauthor{whittle1981arm}, \citeyear{whittle1981arm}). To give a concrete example, suppose that all
treatments start with a common uniform prior, then all initial states
are equal to $\mathbf{x}_{k,0}=(1,1)$ with a corresponding Gittins
index value of $0.8699$ for all of them. Yet, if a treatment $k$ has a
beta prior with parameters $(1,2)$ and another treatment $k'$ has a
prior with parameters $(2,1)$, their respective initial states are
$\mathbf{x}_{k,0}=(1,2)$ and $\mathbf{x}_{k',0}=(2,1)$, and their
associated index values respectively are $0.7005$, $0.9102$. The same
reasoning applies for the case in which priors combine with data so as
to have $\mathbf{x}_{k,1}=(1,2)$ and $\mathbf{x}_{k',1}=(2,1)$.

The underlined values in Table~\ref{tab:Gittinstable} describe situations
in which the \emph{learning} element plays a key role.
Consider two treatments with the same
posterior mean of success $2/4=4/8=1/2$. According to the indices
denoted by the single line, the treatment with the smallest number of
observations is preferred: $0.7844>0.6952$. Moreover, consider the case
in which the posterior means of success suggest the superiority of one
over the other: $2/5=0.4 < 6/12 = 0.5$, yet their indices denoted by
the double-underline suggest the opposite, $0.6726>0.6504$, again
prioritizing the least observed population.

\citet{gittins1992learning} define the \emph{learning} component of the
index as the difference between the index value and the expected
immediate reward, which for the general Bayesian Bernoulli MABP is
given by $\frac{s_{k,0}+s_{k,t}}{s_{k,0}+f_{k,0}+s_{k,t}+f_{k,t}}$.
This posterior probability is the current belief that a treatment $k$
is successful and it
can be used for making patient allocation decisions in a myopic way,
that is, exploiting the available information without taking into
account the possible future learning.
Consider, for instance, the case where $\mathbf{x}_{k,0}=(1,1)$ for all $k$.
In that case, the learning component
before making any treatment allocation decision is thus $(0.8699 -0.5)=0.3699$.
As the number of observations of a bandit increases, the learning part
of the indices decreases.

\section{The Finite Horizon Case: A Restless MABP} \label
{restless MABP}

Of course, clinical trials are not run with infinite resources or
patients. Rather, one usually attempts to recruit the minimum number of
patients to achieve a pre-determined power. Thus, we now consider the
optimization problem defined in \eqref{eq:dop_1} for a finite value of
$T$. Indeed, a solution could in theory be obtained via DP, but it is
impractical in large-scale scenarios for reasons already stated.
Moreover, the Index theorem does not apply to this case, thus, the
Gittins index function as defined for the infinite-horizon variant does
not exist \citep{berry1985bandit}. In the infinite-horizon problem, at
any $t$ there is always an infinite number of possible sample
observations to be drawn from any of the populations. This is no longer
the case in a finite-horizon problem, and the \emph{value} of a
sampling history $(s_{k,t}, f_{k,t})$ is not the same when the sampling
process is about to start than when it is about to end. The
finite-horizon problem analysis is thus more complex, because these
transient effects must be considered for the characterization of the
optimal policy. In what follows we summarize how to derive an index
function analogous to Gittins' rule for the finite-horizon Bayesian
Bernoulli MABP based on an equivalent reformulation of it as an
infinite-horizon \emph{Restless} MABP, as it was done in \citet
{nino2005marginal}. In the equivalent model the information state is
augmented, adding the number of remaining sample observations that can
be drawn from the $K$ populations. Hence, the MCP has the following
modified elements:
\begin{longlist}[b]
\item[(a)] An augmented \emph{state space} $\mathbb{\hat{X}}_k$
given by
the union of the set $\mathbb{X}_{k,t} \times\mathbb{T}$, where
$\mathbb{T}= \{0,1, \dots, T\}$, and an absorbing state $\{ E \}$,
representing the end of the sampling process.
Thus,
$\hat{\mathbf{x}}_{k,t}=(\mathbf{x}_{k,t}, T-t)$ is a
three-dimensional vector combining the
information on the treatment (prior and observed) and the
number of remaining patients to allocate until the end of the trial.%

\item[(b)] The same as in Section~\ref{sec:2 mabp}.

\item[(c)] A
\emph{transition law} $\mathcal P_k(\hat{\mathbf{x}}_{k,t+1}|\hat
{\mathbf{x}}_{k,t},a_k)$
for every $\hat{\mathbf{x}}_{k,t} \mbox{ such that } 0\le t \le T-1 $:
%
\begin{eqnarray}
\label{dynamics2} &&\hat{\mathbf{x}}_{k,t+1} \hspace*{-24pt}\nonumber
\\[-4pt]\\[-16pt]
&&\quad= \cases{ \mbox{if
}a_{k,t}=1: &
\cr
\quad\bigl(s_{k,0}+s_{k,t}+1,f_{k,0}+f_{k,t},
T-(t+1) \bigr), \cr
\qquad \mbox{w.p. }
 \displaystyle\frac{ s_{k,t}+s_{k,0}}{s_{k,t}+f_{k,t}+s_{k,0+}f_{k,0}} ,
\cr
\quad\bigl(s_{k,0}+s_{k,t},f_{k,0}+f_{k,t}+1,
T-(t+1) \bigr), \cr
\qquad \mbox{w.p. } \displaystyle\frac{f_{k,t}+f_{k,0}}{s_{k,t}+f_{k,t}+s_{k,0}+f_{k,0}},
\cr
\mbox{if }
a_{k,t}=0 \quad \bigl(\mathbf{x}_{k,t}, T-(t+1) \bigr) ,\cr
 \qquad\mbox{w.p.
} 1, }\hspace*{-24pt}\nonumber
\end{eqnarray}
$\hat{\mathbf{x}}_{k,T}$ and $E$, under both actions, lead to $E$
with probability one.

\item[(d)] The one-period expected rewards and resource consumption
functions are defined as in
\eqref{rewards1} for $t =0,1, \dots, T-1$, while
the states $E$ and $\hat{\mathbf{x}}_{k,T}$
both
yield $0$ reward and work consumption.
\end{longlist}

The objective in the resulting bandit optimization problem is also to
find a discount-optimal policy that maximizes the ETD rewards.

\subsection{Restless MABPs and the Whittle Index}\label{restless}
In this equivalent version 
the horizon is infinite (a~fiction introduced by forcing every arm of
the MABP to remain in state $E$ after the period $T$), nonetheless,
the Index theorem 
does not apply to it because its dynamics
do not fulfil condition \eqref{classic dyn}. The inclusion of the
number of remaining observations to allocate as a state variable causes
inactive arms to evolve regardless of the selected action, and this
particular feature makes the augmented MABP \emph{restless}.

In the seminal work by \citet{whittle1988rba}, this particular
extension to the MABP dynamics was first proposed and the name \emph
{restless} was introduced to refer to this class of problems. Whittle
deployed a Lagrangian relaxation and decomposition approach to derive
an index function, analogous to the one Gittins had proposed to solve
the \emph{classic} case, which has become known as the Whittle index.

One of the main implications of Whittle's work is the realization that
the existence of such an index function is not guaranteed for every
\emph{restless} MABP. Moreover, even in those cases in which 
it exists, the index rule does not necessarily recover the optimal
solution to the original MABP (as it does in the \emph{classic} case),
being thus a heuristic rule.
Whittle further
conjectured that the index policy for the restless variant enjoys a
form of asymptotic optimality (in terms of the ETD rewards achieved), a
property later established by \citet{weber1990index} under certain
conditions. Typically, the resulting heuristic has been found to be
nearly optimal in various 
models.

\subsection{Indexability of Finite-Horizon Classic MABP}\label
{fh index}

In general, establishing the existence of an index function for a \emph
{restless} MABP (i.e., showing its \emph{indexability}) and computing
it is a tedious task. In some cases, the sufficient indexability
conditions (SIC) introduced by \citet{nino2001restless} can be applied
for both purposes.

The restless bandit reformulation of finite-horizon \emph{classic}
MABPs, as defined in Section~\ref{sec:2 mabp}, is always \emph{indexable}.
Such a property can either be shown by means of the SIC approach
or simply using the seminal result in \citet{bellman1956problem}, by
which the monotonicity of the optimal policies can be ensured, allowing
to focus attention on a nested family of stopping times.

Moreover, the fact that in this
restless MABP reformulation
the part of the augmented state that continues to evolve under
$a_{k,t}=0$, that is, $T-t$, does so in the exact same way that under
$a_{k,t}=1$ allows computation of the Whittle index as a modified
version of the Gittins index, in which the search of the optimal
stopping time in \eqref{Gittins} is truncated to be less than or equal
to the number of remaining observations to allocate (at each decision
period) (see Proposition~3.1 in \citeauthor{nino2011computing}, \citeyear{nino2011computing}). Hence,
the Whittle index for the finite-horizon Bayesian Bernoulli MABP is
%
\begin{eqnarray}
\label{Whittle} &&\mathcal W_k(\hat{\mathbf{x}}_{k,t})= \sup
_{1\le\tau\le T-t}\frac{\Ex
_{\hat{\mathbf{X}}_{k,t}=\hat{\mathbf{x}}_{k,t}} \sum_{i=0}^{\tau-1} \mathcal R(\hat{\mathbf{X}}_{k,t+i},1) d^i }{\Ex_{\hat
{\mathbf{X}}_{k,t}=\hat{\mathbf{x}}_{k,t}} \sum_{i=0}^{\tau
-1}\mathcal C(\hat{\mathbf{X}}_{k,t+i},1) d^i }, \nonumber
\\[-10pt]\\[-6pt]
&&\hspace*{114pt}\quad\mbox{for } \hat {
\mathbf{x}}_{k,t} \in\mathbb{\hat{\mathbf{X}}}_k \setminus
\{E, \hat {\mathbf{x}}_{k,T} \},\hspace*{-114pt}\nonumber
\end{eqnarray}
where the expectation is computed with respect to the corresponding
Markovian (\emph{active}) transition law $\mathcal{P}_k(\hat{\mathbf
{x}}_{k,t+1}|\hat{\mathbf{x}}_{k,t},1)$ and $\tau$ is a stopping time.

Table~\ref{tab:Whittletable}, Table~\ref{tab:Whittletable1} and Table~\ref{tab:Whittletable2} include some values of the Whittle indices for
instances in which, as before, the initial prior is uniform for all the
arms and the index precision is of 4 digits, but the discount factor is
$d=1$, the sampling horizon is set to be
$T=180$, and the number of remaining observations is respectively
allowed to be $T-t=80$, $T-t=40$ and $T-t=1$.
Again, the Whittle index rule
assigns a number from these tables to every treatment, based on the
values of $s_{k,0}+s_{k,t}$ and $f_{k,0}+f_{k,t}$ and on the number of
remaining periods $T-t$, and then prioritizes sampling the one with the
highest value.
%
\begin{table}
\caption{The Whittle index values for an information vector of
$s_{0}+s_{t}$ successes and $f_0+f_t$ failures, $T-t=80$, $d=1$ and
where the size of the trial is $T=180$}
\label{tab:Whittletable}
\begin{tabular*}{\columnwidth}{@{\extracolsep{\fill}}lcccccc@{}}
\hline
$\bolds{f/s}$& \textbf{1} &
\textbf{2} & \textbf{3} & \textbf{4} & \textbf{5} & \textbf{6} \\
\hline
{1} & $0.8558$ & $0.9002$ & $0.9204$& $0.9326$ & $0.9409$ & $0.9471$ \\
{2} &$ 0.6803$&$ \underline{0.7689}$ & $0.8140$ & $0.8423$ &$0.8621$ &
$0.8769$\\
{3} & $0.5463$ & $\underline{\underline{0.6552}}$ &$0.7158$ &$0.7565$
&$0.7855$ &$0.8077$\\[3pt]
{4} & $0.4503$ & $0.5630$ & $0.6335$ & $\underline{0.6812}$ & $0.7167$ &
$0.7444$ \\[1pt]
{5} & $0.3786$ &$0.4923$ &$0.5642$ &$0.6169$ &$0.6565$ &$0.6876$\\
{6} & $0.3247$ & $0.4348 $& $0.5073$ & $0.6040$ & $0.6040$ &$ \underline
{\underline{0.6380}}$\\
\hline
\end{tabular*}
 \end{table}

\begin{table}[b]
\caption{The Whittle index at $T-t=40$}
\label{tab:Whittletable1}
\begin{tabular*}{\columnwidth}{@{\extracolsep{\fill}}lcccccc@{}}
\hline
$\bolds{f/s}$& \textbf{1} &
\textbf{2} & \textbf{3} & \textbf{4} & \textbf{5} & \textbf{6} \\
\hline
{1} & $0.8107$ & $0.8698$ & $0.8969$ & $0.9132$ & $0.9244$ & $0.9326$ \\
{2} & $0.6199$ & $\underline{0.7239}$& $0.7778$ & $0.8120$& $0.8360$&
$0.8539$ \\[1pt]
{3} & $0.4877$ &$\underline{\underline{0.6067}}$ & $0.6753$ & $0.7214$ &
$0.7546$ & $0.7802$\\[3pt]
{4} & $0.3955$ & $0.5157$ & $0.5920$ & $\underline{0.6447}$ & $0.6837 $&
$0.7147$ \\[1pt]
{5} & $0.3297 $& $0.4476$ & $0.5231 $& $0.5802 $& $0.6233$ & $0.6573$ \\
{6} & $0.2805 $& $0.3929 $& $0.4690$ & $0.5254 $& $0.571 $& $\underline
{\underline{0.6075}}$ \\
\hline
\end{tabular*}
\end{table}

It follows from the above tables
that the \emph{learning} element of this index
decreases as $T-t$ decreases. In the limit, when $T-t=1$ the Whittle
index is exactly the posterior mean of success (which corresponds to
the \emph{myopic} allocation rule that results from using current
belief as an index). On the contrary, as $T-t \to\infty$, the Whittle
index tends to approximate the Gittins index. Hence, for a given
information vector, the relative importance of exploring (or \emph
{learning}) vs. exploiting (or being \emph{myopic}) varies
significantly over time
in a finite-horizon problem as opposed to the infinite-horizon case in
which this balance remains constant in time depending solely on the
sampling history. Notice that the
computational cost of a single Whittle index table is, at most, the
same as for a Gittins index one; however, solving a finite horizon MABP
using the Whittle rule has significantly higher computational cost than
the infinite-horizon case, because the Whittle indices must
be computed at every time point $t$.

\begin{table}
\caption{The Whittle index at $T-t=1$}
\label{tab:Whittletable2}
\begin{tabular*}{\columnwidth}{@{\extracolsep{\fill}}lcccccc@{}}
\hline
$\bolds{f/s}$& \textbf{1} &
\textbf{2} & \textbf{3} & \textbf{4} & \textbf{5} & \textbf{6} \\
\hline
 {1} &$ 0.5000$ & $0.6667$ & $0.7500$ & $0.8000$ & $0.8333$ & $0.8571$ \\
{2} & $0.3333$ & $\underline{0.5000}$ & $0.6000 $& $0.6667 $& $0.7143$ &
$0.7500$ \\[1pt]
{3} & $0.2500$ &$ \underline{\underline{0.4000}}$ & $0.5000 $& $0.5714$
& $0.6250$ & $0.6667$\\[3pt]
{4} & $0.2000$ & $0.3333$ & $0.4286$ & $\underline{0.5000}$ & $0.5556$ &
$0.6000$ \\[1pt]
{5} &$ 0.1667$ & $0.2857$ & $0.3750$ & $0.4444 $& $0.5000$ & $0.5455$ \\
{6} & $0.1429$ & $0.2500$ & $0.3333$ & $0.4000$ & $0.4545$ & $\underline
{\underline{0.5000}}$ \\
\hline
\end{tabular*}
\end{table}

\begin{figure}[b]

\includegraphics{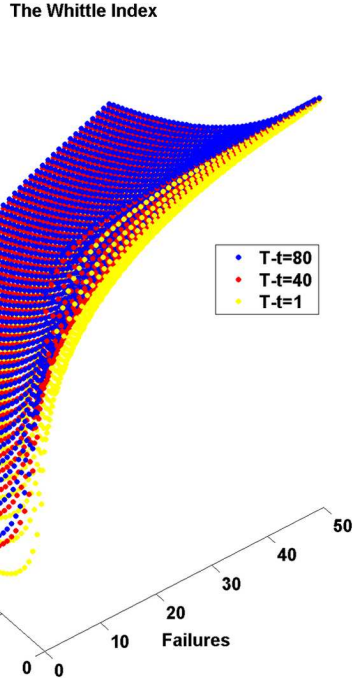}

\caption{The (approximate) Whittle index values for an information
vector of $s_{0}+s_{t}$ successes and
$f_0+f_t$ failures, plotted for $T-t \in\{1,40,80\}$ with $d = 1$ and
$T=180$.}
\label{fig3}
\end{figure}

This evolution of the learning vs. earning trade-off is depicted graphically
in Figure~\ref{fig3} and
causes the decisions in each of the highlighted situations of Table~\ref{tab:Gittinstable} to change over time when considered for a
finite-horizon problem. In Table~\ref{tab:Whittletable}
with $T-t=80$
both decisions coincide with the ones described for Table~\ref{tab:Gittinstable}, while in Table~\ref{tab:Whittletable1}, in which $T-t=40$,
the decision for the second example has changed,
and in Table~\ref{tab:Whittletable2}, in which $T-t=1$,
the decisions in both cases are different.

%

\section{Simulation Study}\label{suitability}

In this section we evaluate the performance of a range of patient
allocation rules in a clinical trial context, including the
bandit-based solutions of Section~\ref{classic MABP} and Section~\ref{restless MABP}. We focus on the following: statistical power $(1-\beta)$; type-I
error rate ($\alpha$); expected proportion of patients in the trial
assigned to the best treatment ($p^*$); expected number of patient
successes (ENS); and, for the two-arm case, bias in the maximum
likelihood estimate of treatment effect associated with each decision
rule. Specifically, we investigate the following patient allocation procedures:
\begin{itemize}
\item\emph{Fixed Randomized design (FR)}: uses an equal, fixed
probability to allocate patients to each arm throughout the trial.
\item\emph{Current Belief (CB)}: allocates each patient to the
treatment with the highest mean posterior probability of success.
\item\emph{Thompson Sampling (TS)}: randomizes each patient to a
treatment $k$ with a
probability that is proportional to the posterior probability that
treatment $k$ is the best given the data. In the simulations we shall
use the allocation probabilities defined as
%
\begin{eqnarray}
\label{thompsonsampling} \pi_{k,t}&=&\mathcal{P}(a_{k,t}=1|
\mathbf{x}_{k,t})\nonumber
\\[-10pt]\\[-10pt]
&=&\frac{\mathcal
{P}(\max_{i} p_i=p_k| \mathbf{x}_{k,t})^c}{\sum_{k=1}^K\mathcal{P}(\max_{i} p_i=p_k| \mathbf{x}_{k,t})^c},\nonumber
\end{eqnarray}
where $c$ is a tuning parameter defined as $\frac{t}{2T}$, and $t$ and
$T$ are the current and maximum sample size respectively. See, for
example, \citet{thall2007practical}.
\item\emph{Gittins index (GI)} and \emph{Whittle index (WI)}:
respectively use the corresponding index functions defined by formulae
\eqref{GittinsB} and \eqref{Whittle}.
\item\emph{Upper Confidence Bound index (UCB)}: developed by \citet
{auer2002finite}, takes into account not only the posterior mean but
also its variability by
allocating the next patient to the treatment with the highest value of an
index, calculated as follows: $\frac
{s_{k,0}+s_{k,t}}{s_{k,0}+f_{k,0}+s_{k,t}+f_{k,t}} + \sqrt{\frac{2 \log
{t}}{s_{k,0}+f_{k,0}+s_{k,t}+f_{k,t}} }$.
\end{itemize}

\subsubsection*{Semi-Randomized (Asymptotically Optimal) Bandit Approaches}

In addition, we consider a randomized class of index-based bandit
patient allocation procedures based on a simple modification %
first suggested in \citet{bather1981randomized}. The key idea is to add
small perturbations to the index value corresponding to the observed
data at each stage, obtaining a new set of indices in which the
(deterministic) index-based part captures the importance of the \emph
{exploitation} based on the accumulated information and the (random)
perturbation part captures the \emph{learning} element. Formally, these
rules are defined as follows:
%
\begin{eqnarray}
\label{randomisedrules} &&I(s_{k,0}+s_{k,t},f_{k,0}+f_{k,t})
\nonumber
\\[-10pt]\\[-10pt]
&&\quad{}+ Z_t* \lambda (s_{k,0}+s_{k,t}+f_{k,t}+f_{k,0}),\nonumber
\end{eqnarray}
where $I(s_{k,0}+s_{k,t},f_{k,0}+f_{k,t})$ is the index value
associated to the prior and observed data on arm $k$ by time $t$, $Z_t$
is an i.i.d. positive and unbounded random variable, and
$\lambda(s_{k,0}+s_{k,t}+f_{k,t}+f_{k,0})$ is a sequence of strictly
positive constants tending to $0$ as $s_{k,0}+s_{k,t}+f_{k,t}+f_{k,0}$
tends to $\infty$. The interest in this class of rules is due to their
asymptotic optimality, that is, property \eqref{regret} discussed in
Section~\ref{sec:2 mabp}, specifically on assessing how their performance
compares to the index rules that are optimal (or nearly optimal) in
terms of   the ETD objective \eqref{eq:dop_1}. Notice that rules
defined by \eqref{randomisedrules} have a decreasing, though strictly
positive, probability of allocating patients to every arm at any point
of the trial. In other words, rules \eqref{randomisedrules} are such
that most of the patients are allocated sequentially to the current
best arm (according to the criteria given by the index value), while
some patients are allocated all the other of the treatment arms.

For the simulations included in this paper we let $Z_t(K)$ be an
exponential random variable with parameter $\frac{1}{K}$; $\lambda
(s_{k,0}+s_{k,t}+f_{k,t}+f_{k,0}) = \frac
{K}{s_{k,0}+s_{k,t}+f_{k,t}+f_{k,0}}$ and define two additional approaches:
\begin{itemize}
\item \emph{Randomized Belief index (RBI) design}: makes the sampling
decisions between the populations based on an index computed setting
$I(s_{k,0}+s_{k,t},f_{k,t}+f_{k,0})=\frac
{s_{k,0}+s_{k,t}}{s_{k,0}+f_{k,0}+s_{k,t}+f_{k,t}}$ in
 \eqref{randomisedrules}.
\item \emph{Randomized Gittins index (RGI) design}: first suggested in
\citet{glazebrook1980randomized}, makes the sampling decisions between
the populations based on the index computed setting
$I(s_{k,0}+s_{k,t},f_{k,t}+f_{k,0})=\mathcal G(s_{k,0}+s_{k,t},
f_{k,t}+f_{k,0})$ in \eqref{randomisedrules}.
\end{itemize}

For every design, ties are broken at random and in every simulated
scenario we let $\mathbf{x}_{k,0}=( s_{k,0},f_{k,0})=(1,1)$ for all $k$.

\subsubsection*{Design Scenarios}

We implement all of the above methods in several $K$-arm trial design
settings. In each case, trials are made up of $K-1$ experimental
treatments and one control treatment. The control group (and its
associated quantities) is always denoted by the subscript $0$ and the
experimental treatment groups by $1,\ldots,K-1$. We first consider the
case $K=2$. To compare the two treatments, we consider the following
hypothesis: $H_0: p_0 \ge p_1$, with the type-I error rate calculated
at $p_0=p_1=0.3$ and the power to reject $H_0$ calculated at $H_{1}:
p_0 =0.3 ; p_1=0.5$. We set the size of the trial to be $T=148$ to
ensure that  FR  will attain at least $80\%$ power when rejecting $H_0$
with a one-sided $5\%$ type-I error rate. We then evaluate the
performances of these designs by simulating $10^4$ repetitions of the
trials under each hypothesis and comparing the resulting operating
characteristics of the trials. Hypothesis testing is performed
using a normal cutoff value (when appropriate) and using an adjusted
Fisher's exact test for comparing two binomial distributions, where the
adjustment chooses the cutoff value to achieve a 5\% type-I error. %

%
\begin{table*}
\tabcolsep=0pt
\caption{Comparison of different two-arm trial designs of size $T=
148$. $F_a$: Fisher's adjusted test; $\alpha$: type-I error; $1-\beta$:
power; $p^{*}$: expected proportion of patients in the trial assigned
to the best treatment; ENS: expected number of patient successes;
UB: upper bound}
\label{tab:simulations1}
\begin{tabular*}{\textwidth}{@{\extracolsep{\fill}}lccccccc@{}}
\hline
& \multirow{2}{25pt}{\textbf{Crit.} \textbf{value}}
&
\multicolumn{3}{c}{$\bolds{H_0: p_0=p_1=0.3}$}
&
\multicolumn{3}{c}{$\bolds{H_1: p_0=0.3,  p_1=0.5}$}
\\
\ccline{3-5,6-8}
&  & \multicolumn{1}{c}{$\bolds{\alpha}$}
&\multicolumn{1}{c}{$\bolds{p^*}$ \textbf{(s.e.)}} & \textbf{ENS} \textbf{(s.e.)} &
\multicolumn{1}{c}{ $\bolds{1-\beta}$} & \multicolumn{1}{c}{$\bolds{p^*}$
\textbf{(s.e.)}}& \textbf{ENS (s.e.)} \\
\hline
{FR} & 1.645 & 0.052 & 0.500 (0.04) & 44.34 (5.62) & 0.809 & 0.501
(0.04) & 59.17 (6.03)\\
{TS} & 1.645 & 0.066 & 0.499 (0.10) & 44.39 (5.58) & 0.795 & 0.685
(0.09) & 64.85 (6.62) \\
{UCB} &1.645 & 0.062 & 0.499 (0.10) & 44.30 (5.60) & 0.799 & 0.721
(0.07) & 66.03 (6.57) \\
{RBI} &1.645 & 0.067 & 0.502 (0.14) & 44.40 (5.57) & 0.763 & 0.737
(0.07) & 66.43 (6.54) \\
{RGI} & 1.645 & 0.063 & 0.500 (0.11) & 44.40 (5.61) & 0.785 &
0.705 (0.07) & 65.46 (6.40) \\%
{CB} &\emph{$F_a$} & 0.046 & 0.528 (0.44) & 44.34 (5.55) & 0.228 &
0.782 (0.35) & 67.75 (12.0) \\
{WI} &\emph{$F_a$} & 0.048 & 0.499 (0.35) & 44.37 (5.59) & 0.282 &
0.878 (0.18) & 70.73 (8.16) \\
{GI} &\emph{$F_a$} & 0.053 & 0.501 (0.26) & 44.41 (5.58) & 0.364 &
0.862 (0.11) & 70.21 (7.11) \\[6pt]
 {UB}  & & & & 44.40 (0.00) &&
 1 & 74.00 (0.00) \\
 \hline
\end{tabular*}
\end{table*}

For the $K$-arm design settings we shall consider the following hypothesis:
$H_0: p_0\ge p_i$ for $i=1, \dots,K-1$ with the family-wise error rate
calculated at $p_0=p_1=\cdots=p_{K-1}=0.3$.
We use the Bonferroni correction method to account for multiple testing
and therefore ensure that the family-wise error rate is less than or
equal to 5\%, that is, all hypotheses whose $p$-values $p_{k}$ are such
that $p_{k}<\frac{\alpha}{K-1}$ are rejected. Additionally, when there
are multiple experimental treatments, we shall define the statistical
power as the probability of the trial ending with the conclusion that a
truly effective treatment is effective.

\subsection{Two-Arm Trial Setting Simulations}

Table~\ref{tab:simulations1} shows the results for $K=2$ under both
hypotheses and for each proposed allocation rule.
The randomized and semi-randomized response-adaptive procedures (i.e.,
TS, UCB, RBI and RGI) exhibit a slightly inferior power level than a FR
design; however, they have an advantage in terms of ENS
over a FR design.
On the other hand, the three deterministic index-based approaches
(i.e., CB, WI and GI) have the best performance in terms of ENS, yet
result in power values which are far below the required values. In the
most extreme case, for the CB and WI rules, the power is approximately
3.5 times smaller than with a FR design.

Adaptive rules have their power reduced because they induce correlation
among treatment assignments; however, for the deterministic index
policies this effect is the most severe because they permanently skew
treatment allocation toward a treatment as soon as one exhibits a
certain advantage over the other arms.

To illustrate the above point, let $n_0$ and $n_1$ be the number of
patients allocated to treatment 0 and 1 respectively, then for the
results in Table~\ref{tab:simulations1} it holds that $E^{\mathrm{CB}}(n_0)=
31.60$, $E^{\mathrm{CB}}(n_1)= 116.40$, $E^{\mathrm{WI}}(n_0)=16.49$,
$E^{\mathrm{WI}}(n_1)=131.51$ and $E^{\mathrm{GI}}(n_0)= 19.06$, $E^{\mathrm{GI}}(n_1)= 128.94$.
Moreover, this implies that the required ``superiority'' does not need
to be a statistically significant difference of the size included in
the alternative hypothesis as suggested by the following values:
$E^{\mathrm{CB}}_k(\mathbf{s}/\mathbf{n})= [0.1437  ; 0.4208]$,
$V^{\mathrm{CB}}_k(\mathbf{s}/\mathbf{n})= [0.1528  ; 0.1831]$,
$E^{\mathrm{WI}}_k(\mathbf{s}/\mathbf{n})= [0.1976  ;\break  0.4860]$,
$V^{\mathrm{WI}}_k(\mathbf{s}/\mathbf{n})= [0.1470  ; 0.08875]$,
$E^{\mathrm{GI}}_k(\mathbf{s}/\mathbf{n})= [0.2283  ; 0.4959]$ and
$V^{\mathrm{GI}}_k(\mathbf{s}/\mathbf{n})= [ 0.1271  ; 0.0538]$.

The results in Table~\ref{tab:simulations1} illustrate the natural
tension between the two opposing goals of maximizing the statistical
power to detect significant treatment effects (using FR) and maximizing
the health of the patients in the trial (using GI). The optimality
property inherent in the GI design produces an average gain in
successfully treated patients of 11 (an improvement of $18.62\%$ over
the FR design). This is only 4 fewer patients on average than the
theoretical upper bound (calculated as $T\times p_{1} = 74$) achievable
if all patients were assigned to the best treatment from the start.
It is worth noting that the asymptotically optimal index approaches
[w.r.t. \eqref{regret}] improve on the statistical power of the index
designs (around  76\%--78\% for a  5\%  type-I error rate) at the
expense of attaining an inferior value of ENS (around 5 fewer successes
on average compared to the bandit-based rules). Yet, these rules
significantly improve on the value of ENS attained by a FR design,
naturally striking a better balance in the patient health/power trade-off.

\begin{figure*}

\includegraphics{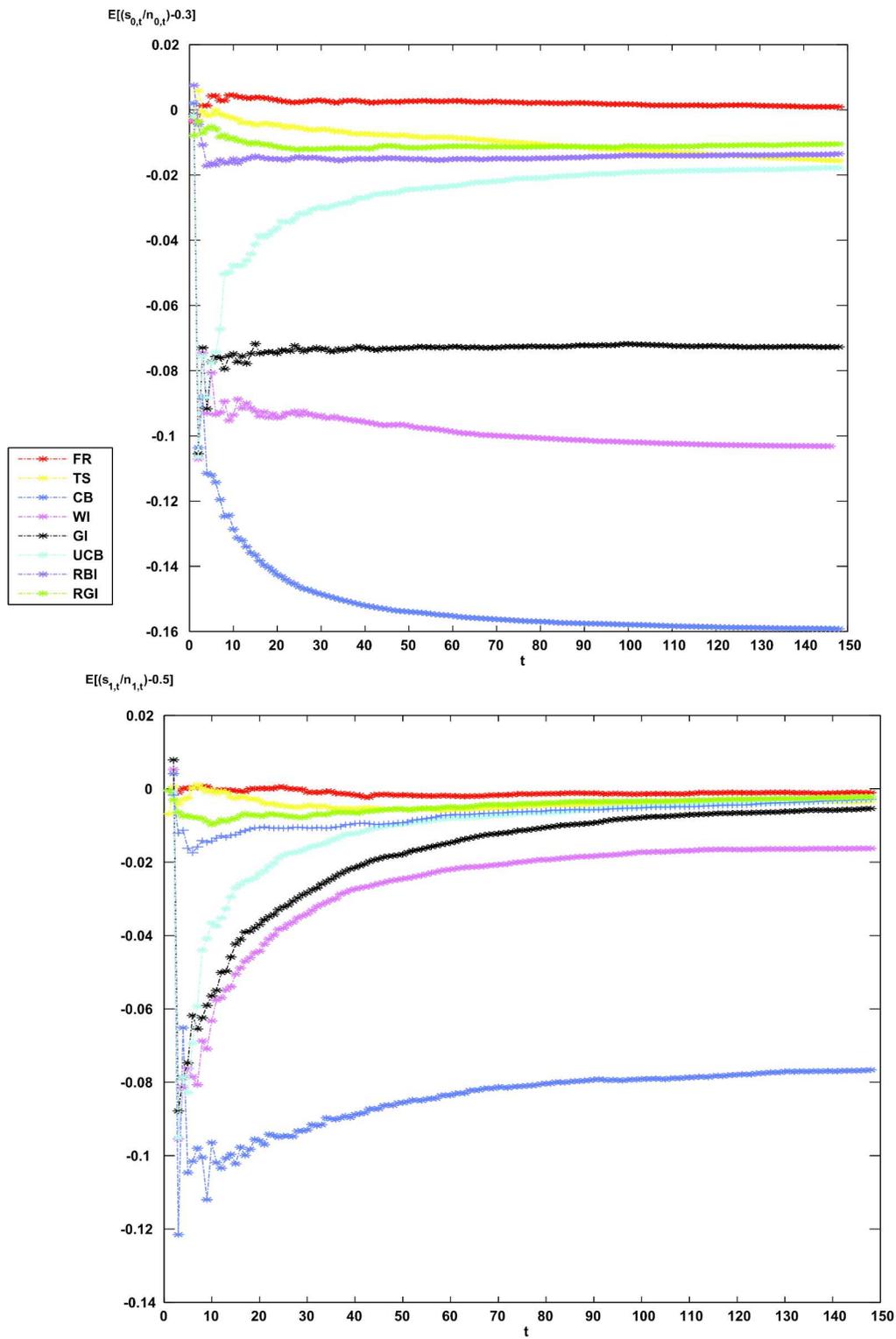}

\caption{Top: The bias in the control treatment estimate as a function
of the number of allocated patients under $H_{1}$. Bottom: The bias in
the experimental treatment estimate under $H_{1}$.}
\label{fig1}
\end{figure*}
From Table~\ref{tab:simulations1} one can see that the three index-based
rules significantly improve on the average number of successes in the
trial by increasing the allocation toward the superior treatment based
on the observed data. This acts to reduce the power to detect
significant treatment effect. Another factor at play is bias:
index-based rules\vadjust{\goodbreak} induce a negative bias in the treatment effect
estimates of each arm, the magnitude of this bias is largest for
inferior treatments (for which less patients are assigned to than
superior treatments). When the control is inferior to the experimental
treatment, this induces a positive bias in the estimated benefit of the
experimental treatment over the control. 
This is shown in Figure~\ref{fig1}. A heuristic explanation for this is as
follows. The index-based rules select a ``superior'' treatment before
the trial is over based on the accumulated data. This implies that
if a treatment performs worse than its true average, that is, worse for
a certain number of consecutive patients, then the treatment will not
be assigned further patients. The treatment's estimate then has no
chance to regress up toward the true value. Conversely, if a treatment
performs better than its true average, the index-based rules all assign
further patients to receive it, and its estimate then has the scope to
regress down toward its true value. This negative bias of the
unselected arms is observed for all dynamic allocation rules, and is
the most extreme for the CB method.

\begin{table*}
\tabcolsep=0pt
\caption{Comparison of different four-arm trial designs of size $T=
423$. $F_a$: Fisher's adjusted test; $\alpha$: family-wise type-I
error; $1-\beta$: power; $p^{*}$: expected proportion of patients in
the trial assigned to the best treatment; ENS: expected number of
patient successes;  UB: upper bound}
\label{tab:simulations2}
\begin{tabular*}{\textwidth}{@{\extracolsep{\fill}}lcccccccc@{}}
\hline
& \multirow{2}{25pt}{\textbf{Crit.} \textbf{value}}
&
\multicolumn{3}{c}{$\bolds{H_0: p_0=p_i=0.3}$ \textbf{for} $\bolds{i=1,\dots,3}$}
&
\multicolumn{3}{c}{$\bolds{H_1: p_0=p_i=0.3}$\textbf{,} $\bolds{i=1,2}$\textbf{,} $\bolds{p_3=0.5}$}
\\
\ccline{3-5,6-8}
&  & \multicolumn{1}{c}{$\bolds{\alpha}$}
&\multicolumn{1}{c}{$\bolds{p^*}$ \textbf{(s.e.)}} & \textbf{ENS} \textbf{(s.e.)} &
\multicolumn{1}{c}{ $\bolds{(1-\beta)}$} & \multicolumn{1}{c}{$\bolds{p^*}$
\textbf{(s.e.)}}& \textbf{ENS (s.e.)} \\
\hline
{FR} & \emph{2.128} & 0.047 & 0.250 (0.02) & 126.86 (9.41) & 0.814
& 0.250 (0.02) & 148.03 (9.77) \\
{TS} &\emph{2.128} & 0.056 & 0.251 (0.07) & 126.93 (9.47) & 0.884
& 0.529 (0.09) & 172.15 (13.0) \\
{UCB} &\emph{2.128} &0.055 & 0.251 (0.06) & 126.97 (9.41) & 0.877
& 0.526 (0.07) & 171.70 (11.9) \\
{RBI} &\emph{2.128} & 0.049 & 0.250 (0.03) & 126.77 (9.40) & 0.846
& 0.368 (0.04) & 158.34 (10.4) \\
{RGI} & \emph{2.128} & 0.046 & 0.250 (0.03) & 126.80 (9.36) &
0.847 & 0.358 (0.03) & 157.26 (10.3)\\
{CB} &\emph{$F_a$} & 0.047 & 0.269 (0.39) & 126.89 (9.61) & 0.213
& 0.677 (0.41) & 184.87 (36.8) \\
{GI} &\emph{$F_a$} & 0.048 & 0.248 (0.18) & 126.68 (9.40) & 0.428
& 0.831 (0.10) & 198.25 (13.7)\\
{CG} & \emph{2.128}& 0.034 & 0.250 (0.02) & 127.16 (9.46) & 0.925
& 0.640 (0.08) & 182.10 (12.3)
\\[6pt]
 UB  & & && 126.90 (0.00) &
 &1 & 211.50 (0.00) \\
\hline
\end{tabular*}
\end{table*}

The final observation refers to the fact that although all the
index-based rules fail to achieve the required level of power to detect
the true superior treatment, they tend to correctly skew patient
allocation toward the best treatment within the trial, when it exists.
For the simulation reported in Table~\ref{tab:simulations1} we have
computed the probability that each rule makes the wrong choice (i.e.,
stops allocating patients to the experimental treatment). These values
are as follows:
$0.1730$, $0.0307$, $0.0035$ for the CB, WI and GI methods respectively.

\subsection{Multi-Arm Trial Setting} \label{1arm}

We now present results for a $K=4$ setting. First, we consider the case
of a trial with $T=423$ patients. As before, we set the size of the
trial to ensure that a \emph{FR} design
results in at least $80\%$ power to detect an effective treatment for a
family-wise error rate of less than $5\%$. Results for this case are
depicted in Table~\ref{tab:simulations2}. The Whittle index approach is
omitted because for $T$ roughly larger than 150 its performance is near
identical to that attained by the Gittins index but with a
significantly higher computational cost.

In this setting, the randomized and semi-\break randomized adaptive rules
(i.e., TS, UCB, RBI, RGI)
exhibit an advantage over a FR both in the achieved power and in ENS.
The reason for that is that
these rules continue to allocate patients to all arms while they skew
allocation to the best performing arm, hence, ensuring that by the end
of the design
the control arm will have a similar number of observations than with FR
while the best arm will have a larger number. Among these rules, TS and
UCB exhibit the best balance between power-ENS which achieve the $80\%$
power increasing ENS in approximately 23 over a FR design.
The deterministic index-based rules CB and GI increase this advantage
in ENS over a FR design by roughly 36 and 50, respectively. However, a
severe reduction is again observed in the power values of these designs.
On the other hand, the probability that each of these rules makes a
wrong choice (i.e., it does not skew the allocation toward the best
experimental treatment) is $0.2691$ and $0.0051$, respectively, for the
CB and GI.

\subsection{The Controlled Gittins Index Approach} \label{cgittins}

To overcome the severe loss of statistical power of the Gittins index,
we introduce, for the multi-arm trial setting only, a composite design
in which the allocation to the control treatment is done in such a way
that one in every $K$ patients is allocated to the control group while
the allocation of the remaining patients among the experimental
treatments is done using the Gittins index rule. We refer to this
design as the \emph{controlled Gittins} (CG) approach.

Based on the simulation results, CG manages to solve the trade-off
quite successfully, in the sense that it achieves more than $80\%$
power, while it achieves a mean number of successes very close to the
one achieved by the CB rule and with a third of the variability that CB
exhibits in expected number of patient successes.

\begin{table*}
\tabcolsep=0pt
\caption{Comparison of different four-arm trial designs of size $T=
80$. F: Fisher; $\alpha$: type-I error; $1-\beta$: power; $p^{*}$:
expected proportion of patients in the trial assigned to the best
treatment; ENS: expected number of patient successes;  UB: upper bound}
\label{tab:simulations3}
\begin{tabular*}{\textwidth}{@{\extracolsep{\fill}}lccccccc@{}}
\hline
& \multirow{2}{25pt}{\textbf{Crit.} \textbf{value}} & \multicolumn{3}{c}{$\bolds{H_0: p_0=p_i=0.3}$
\textbf{for} $\bolds{i=1,
\dots,3}$} & \multicolumn{3}{c}{$\bolds{H_1: p_k=0.3+0.1\times k}$\textbf{,} $
\bolds{k=0,1,2,3}$} \\
\ccline{3-5,6-8}
&  & \multicolumn{1}{c}{$\bolds{\alpha}$}
&\multicolumn{1}{c}{$\bolds{p^*}$ \textbf{(s.e.)}} & \textbf{ENS} \textbf{(s.e.)} &
\multicolumn{1}{c}{ $\bolds{(1-\beta)}$} & \multicolumn{1}{c}{$\bolds{p^*}$
\textbf{(s.e.)}}& \textbf{ENS (s.e.)} \\
\hline
{FR} & \emph{F} & 0.019 & 0.251 (0.04) & 24.01 (4.07) & 0.300 &
0.250 (0.04)& 35.99 (4.41) \\
{TS} & \emph{F} & 0.013 & 0.250 (0.07) & 24.01 (4.15) & 0.246 &
0.338 (0.08) & 38.34 (4.68) \\
{UCB} & \emph{F} & 0.011 & 0.252 (0.06) & 24.00 (4.12) & 0.218 &
0.362 (0.08) & 38.84 (4.71) \\
{RBI} & \emph{F} & 0.018 & 0.250 (0.03) & 23.97 (4.06) & 0.295 &
0.268 (0.03) & 36.52 (4.41) \\
{RGI} & \emph{F} & 0.017 & 0.250 (0.02) & 24.07 (4.07) & 0.298 &
0.265 (0.03) & 36.45 (4.36) \\
{CB} &$F_a$ & 0.017 & 0.270 (0.30) & 23.98 (4.08) & 0.056 &
0.419 (0.38) & 40.92 (6.89) \\
{WI} &$F_a$ & 0.015 & 0.258 (0.22) & 23.00 (4.14) &0.101 &
0.537 (0.31) & 42.65 (6.02) \\
{GI} &$F_a$ & 0.000 & 0.251 (0.13) & 23.97 (4.11) & 0.002 &
0.492 (0.21) & 41.60 (5.44) \\
{CG} & $F_a$ & 0.015 & 0.253 (0.13) & 24.04 (4.13) & 0.349
& 0.393 (0.16) & 38.29 (4.82) \\[6pt]
 UB  & && & 24.00 (0.00) &
& 1 & 48.00 (0.00) \\
\hline
\end{tabular*}
 \end{table*}

\subsection{Multi-Arm Trial in a Rare Disease Setting} \label{rare}

Finally, we imagine a rare disease setting, where the number of
patients in the trial is a high proportion of all patients with the
condition, but is not enough to guarantee reasonable power to detect a
treatment effect of a meaningful size. In such a context, the idea of
prioritizing patient benefit over hypothesis testing is likely to raise
less controversy than in a common disease context \citep{Wang18122002}.
We therefore simulate a four-arm trial as before but where the size of
the trial is $T=80$. Given that the size of the trial implies a very
small number of observations per arm, Table~\ref{tab:simulations3} only
includes the results of the tests using Fisher's exact test and
Fisher's adjusted exact test (in this case, adjusted to attain the same
type-I error as the other methods). Also, to make the scenario more
general, we have considered that under the alternative hypothesis the
parameters are such that $H_1: p_k=0.3+0.1\times k$, $k=0,1,2,3$.

The FR approach exhibits a $30\%$ power and attains an ENS value of 36.
Table~\ref{tab:simulations3} shows the results attained for each of the
designs considered. Under the alternative hypotheses, the GI and WI
designs achieve an ENS gain over the FR design of 6 patients. Again,
the CG rule exhibits an advantage over FR both in the achieved power
and in the ENS (which in the case of this small population equals the
advantage achieved by TS or UCB).
Its ENS is less than 10 below the theoretical upper bound of 48. An
important feature to highlight is that the Whittle rule does not
significantly differ from the Gittins rule as it could be expected,
given the trial (and hence its horizon) is small. These results
illustrate how the GI and WI start skewing patient allocation toward
the best arm (when it exists) earlier than other adaptive designs,
therefore explaining their advantage in terms of $p^*$ for small $T$
over all of them.

\section{Discussion}\label{Conclusion}

Multi-armed bandit problems have emerged as the archetypal model for
approaching learning problems while addressing the dilemma of
exploration versus exploitation. Although it has long been used as {\it
the} motivating example, they have yet to find any real application in
clinical trials. After reviewing the theory of the Bernoulli MABP
approach, and the Gittins and Whittle indices in particular, we have
attempted to illustrate their utility compared to other methods of
patient allocation in several multi-arm clinical trial contexts.

Our results in Section~\ref{suitability} show that the Gittins and Whittle index-based
allocation methods perform extremely well when judged solely on patient
outcomes, compared to the traditional fixed randomization approach. The
two indexes have distinct theoretical properties, yet in our
simulations any differences in their performances were negligible, with
both designs being close to each other and the best possible scenario
in terms of patient benefit. Since it only needs to be calculated once
before the trial starts, the Gittins index may naturally be preferred.

The Gittins index, therefore, represents an extremely simple---yet near
optimal---rule for allocating patients to treatments within the finite
horizon of a real clinical trial. Furthermore, since the index is
independent of the number of treatments, it can seamlessly incorporate
the addition of new arms in a trial, by balancing the need to learn
about the new treatment with the need to exploit existing knowledge on
others. The issue of adding treatment arms is present in today's
cutting-edge clinical trials. For example, this facet has been built
into the I-SPY 2 trial investigating tumour-specific treatments for
breast cancer from the start \citep{barker2009}. It is also now being
considered in the multi-arm multi-stage STAMPEDE trial into treatments
for prostate cancer as an unplanned protocol amendment, due to a new
agent becoming available (\citeauthor{sydes2009}, \citeyear{sydes2009};
\citeauthor{wason2012some}, \citeyear{wason2012some}).

Gittins indices and analogous optimality results have been derived for
endpoints other than binary. Therefore, the analysis and conclusions of
this work naturally extend to the multinomial distribution \mbox{\citep
{glazebrook1978optimal}}, normally distributed processes
with known variance \citep{jones1970sequential} and with unknown
variance \citep{jones1975search}, and
exponentially distributed populations
(\citeauthor{do1985aspects}, \citeyear{do1985aspects};
\citeauthor{gittins2011multi}, \citeyear{gittins2011multi}).

Unfortunately, the frequentist properties of designs that utilize
index-based rules can certainly be questioned; both the Gittins and
Whittle index approaches required an adjustment of the Fisher's exact
test in order to attain type-I error control,
produced biased estimates and, most importantly, had very low power to
detect a treatment difference at the end of the trial. Since this
latter issue greatly reduces their practical appeal, we proposed a
simple modification that acted to stabilize the numbers of patients
allocated to the control arm. This greatly increased their power while
seemingly avoiding any unwanted type-I error inflation above the
nominal level. This principle is not without precedence, indeed, \citet
{trippa2012} have recently proposed a Bayesian adaptive design in the
oncology setting for which protecting the control group allocation is
also an integral part. Further research is needed to see whether
statistical tests can be developed for bandit-based designs with
well-controlled type-I error rates and also if bias-adjusted estimation
is possible.

There are of course other obvious limitations to the use of index-based
approaches in practice. A patient's response to treatment needs to be
known before the next patient is recruited, since the subsequent
allocation decision depends on it. This will only be true in a small
number of clinical contexts, for example, in early phase trials where
the outcome is quick to evaluate or for trials where the recruitment
rate may be slow (e.g., some rare disease settings). MABPs rely on this
simplifying assumption for the sake of ensuring both tractability and
optimality, and can not claim these special properties without making
additional assumptions (see, e.g., \citeauthor{caro2010indexability}, \citeyear{caro2010indexability}). It
would be interesting to see, however, if index-based approaches could
be successfully applied in the more general settings where patient
outcomes are observed in groups at a finite number of interim analyses,
such as in a multi-arm multi-stage trial
(\citeauthor{magirr2012generalized}, \citeyear{magirr2012generalized};
\citeauthor{wason2012optimal}, \citeyear{wason2012optimal}). Further research is needed to
address this question.

A different limitation to the use of bandit strategies is found in the
fact that the approach leads to deterministic strategies. Randomization
naturally protects designs against many possible sources of bias, for
example, patient drift unbalancing treatment arms \citep{tang2010} or
unscrupulous trial sponsors cherry-picking patients \citep{fda2006}. Of
course, while these are serious concerns, they could also be leveled at
any other deterministic allocation rule, such as play-the-winner.
Further research is needed to introduce randomization to bandit
strategies and also to determine some general conditions under which
arms are selected or dropped when using the index rules.

Further supporting materials for this paper,\break 
including programs to calculate
extended\break tables of the Gittins and Whittle indexes, can be found at
\surl{http://www.mrc-bsu.cam.ac.uk/software/\\miscellaneous-software/}.

\begin{appendix}
\section*{Appendix: Index Computation}\label{index computation}

There is a vast literature on the efficient computation of the Gittins
indices. In \citet{Beale1979}, \citet{varaiya1985extensions} and \citet
{chen1986linear}, among others, algorithms for computing the Gittins
indices for the infinite-horizon \emph{classic} MABP with a finite
state space are provided. The computational cost for all of them (in
terms of its running time as a function of the number of states $N$) is
$N^3+\mathcal{O}(N^2)$.
The algorithm for computing the Gittins indices in such a case
achieving the lowest time complexity, $2/3 N^3+\mathcal{O}(N^2)$, was
provided by \citet{nino20072}.
For MABP with an infinite state space, such as the Bayesian Bernoulli
MABP in Section~\ref{classic MABP}, the indices can be computed using any
of the above algorithms but confining attention to some finite set of
states, which will eventually determine the precision of their calculation.
For the finite-horizon \emph{classic} MABP, as reviewed in Section~\ref{restless MABP}, an efficient exact computation method based on a
recursive adaptive-greedy algorithm is provided in \citet{nino2011computing}.

In what follows we examine in more detail the so-called \emph
{calibration} method for the approximate index computation in the
Bayesian Bernoulli MABP, both for the
infinite- (Gittins index) and finite-horizon case (Whittle index).
There are many reasons for focusing on this approach, not least because
it was the algorithm used for computing the values presented in this paper.
It also sheds light on the interpretation of the resulting index
values, by connecting the Gittins index approach to
the work in \citet{bellman1956problem}, and has long been the preferred
computational method.

\subsection*{The Calibration Method} \label{calib}

\citet{bellman1956problem}
studied an infinite random sampling problem involving two binomial
distributions: one with a known success rate and the other one with an
unknown rate but with a Beta prior.
Bellman's key contribution was to show that the solution to the problem
of determining the sequence of choices that maximize the ETD
number of successes exists, is unique and, moreover, is expressible in
terms of an index function which depends only on the total observed
number of successes $s$ and failures $f$ of the unknown process.

\citet{gijo74} used that result and showed that the optimal rule for an
infinite-horizon MABP can also be expressed in terms of an index
function for each of the $K$ Bernoulli populations and based on their
observed sampling histories $(s,f)$. Such an index function is given by
the value $p \in[0,1]$ for which
the decision maker is indifferent between sampling the next observation
from a population with known success rate $p$ or from an unknown one
with an expected success rate $\frac{s}{s+f}$. The \emph{calibration}
method uses DP to approximate the Gittins index values based on this
idea, as explained in \citet{gittins1979dynamic}, and it can be adapted
to compute the finite-horizon counterpart, as explained in
\citeauthor{berry1985bandit} (\citeyear{berry1985bandit}), Chapter~5.

Specifically, this index computation
method solves, for a grid of $p$ values (the size of which determines
the accuracy of the resulting index values approximations), the
following DP problem:
%
\begin{eqnarray}
\label{eq:top dp ex}&& V_{D,t}^*(s,f,p)\nonumber
\\
 &&\quad= \max\biggl\{p
\frac{ 1-d^{T-t}}{1-d},\nonumber
\\
&& \hphantom{\quad= \max\biggl\{}\frac{s}{s+f} \bigl(1+ d V_{D,(t+1)}^*
(s+1,f,p) \bigr) \nonumber
\\[-10pt]\\[-10pt]
&& \hphantom{\quad= \max\biggl\{\ }{}+
\frac
{f}{s+f} \bigl( d V_{D,(t+1)}^*(s,f+1,p) \bigr)\biggr\},\nonumber
\\
&&\hspace*{134pt}\quad t=0, \dots, T-2,\nonumber\hspace*{-134pt}
\\
&&V_{D,T-1}^*(s,f,p)  =  \max\biggl\{ p , \frac{s}{s+f}\biggr\}.\nonumber
\end{eqnarray}
For the infinite-horizon problem and with $0\le d <1$, the convergence
result allows for the omission of the subscript $t$ in the optimal
value functions in \eqref{eq:top dp ex}, letting
the reward associated to the known arm be $\frac{p}{(1-d)}$.
For obtaining a reasonably good initial approximation of the optimal
value function, the terminal condition on $V_{D,T-1}^*(s,f,p)$ is
solved for some values of $s$ and $f$ such that $s+f=T-1$, and for a
large $T$ and then a backward induction algorithm is applied to yield
an approximate value for $V_{D,0}^*(s,f,p)$. For a fixed $p$ the total
number of arithmetic operations to solve \eqref{eq:top dp ex} is $1/2
(T-1) (T-2)$, which, as stated in Section~\ref{GItheorem}, no longer grows
exponentially in the horizon of truncation $T$ (nor does it grow in the
number of arms of the MABP).

For the finite-horizon variant, the terminal condition is not used for
approximating the initial point of the backward-induction algorithm and
the solution, but for computing the optimal value function exactly. The
resulting number of operations to compute the Whittle index is
basically the same as for the Gittins index,
yet the total computational cost is significantly higher given that the
Whittle indices must be computed and stored for every
possible $t\le T-1$ and $(s,f)$.
However, notice that an important advantage of the Whittle index over
the Gittins index is that the discount factor $d=1$ can be explicitly
considered for the former directly adopting an Expected Total objective
function, by replacing the term $\frac{1-d^{T-t}}{1-d}$ by $T-t$, using
the fact that
\[
\lim_{d \to1}\frac{1-d^{T-t}}{1-d} = \sum
_{i=0}^{T-t-1}d^i.
\]

%
\end{appendix}

\section*{Acknowledgments}
The authors are grateful for the insightful and very useful
comments of the anonymous referee and Associate Editor that significantly
improved
the presentation of this paper.
Supported in part by the UK Medical Research Council (grant numbers
G0800860 and MR/J004979/1)  and the Biometrika Trust.


%

\end{document}